\documentclass[twocolumn, prb, showpacs, aps]{revtex4} 
\usepackage{hyperref}
\usepackage{graphicx}
\usepackage{amsmath}
\usepackage{amsbsy}
\usepackage{mathbbol}
\usepackage{color}
\usepackage{natbib}
\graphicspath{{figures/}}

\newcommand{\scmd}[1]{{#1}}

\begin{document}

\title{Dynamic spin-Hall effect and driven spin helix for linear spin-orbit interactions}

\author{Mathias Duckheim$^1$}
\author{Dmitrii L. Maslov$^2$}
\author{Daniel Loss$^1$}
\affiliation{$^1$Department of Physics, University of Basel, CH-4056
Basel, Switzerland,}

\affiliation{$^2$Department of Physics, University of Florida, Gainesville, FL 32611 -8440, USA}


\date{\today}
%




\begin{abstract}
  We derive boundary conditions for the electrically induced spin
  accumulation in a finite, disordered 2D semiconductor channel.
  While for DC electric fields these boundary conditions select
  spatially constant spin profiles equivalent to a vanishing spin-Hall
  effect, we show that an in-plane ac electric field results in a
  non-zero ac spin-Hall effect, i.e., it generates a spatially
  non-uniform out-of-plane polarization even for linear intrinsic
  spin-orbit interactions. Analyzing different geometries in [001] and
  [110]-grown quantum wells, we find that although this out-of-plane
  polarization is typically confined to within a few spin-orbit
  lengths from the channel edges, it is also possible to generate
  spatially oscillating spin profiles which extend over the whole
  channel. The latter is due to the excitation of a driven spin-helix
  mode in the transverse direction of the channel. We show that while
  finite frequencies suppress this mode, it can be amplified by a
  magnetic field tuned to resonance with the frequency of the electric
  field.  In this case, finite size effects at equal strengths of
  Rashba- and Dresselhaus SOI lead to an enhancement of the magnitude
  of this helix mode.
  We comment on the relation between spin currents and boundary
  conditions.
 \end{abstract}
\maketitle

\section{Introduction}


Electron systems with spin-orbit interaction show a variety of
spin-electric effects arising from the coupling between (orbital)
charge and spin degrees of freedom.  The most prominent examples are
the spin-Hall effect \cite{Dyakonov1971, Kato2004b, Sih2005,Engel2007a}
and current induced spin polarization \cite{Levitov1985,
  Edelstein1990, Kato2005}, both of which have received substantial
interest due to their potential to generate and control spin
polarization with electric fields. This type of electrical control is
a prerequisite for integrating spin effects into standard lithographic
semiconductor structures and, ultimately, utilizing the spin degree of
freedom as a carrier of information.\cite{spintronics}

The spin-Hall effect (SHE) manifests itself experimentally
\cite{Kato2004b,Wunderlich2005,Sih2005} as current induced spin
polarization (CISP) at the edges of a Hall-bar (in the absence of a
magnetic field). Initial theoretical studies of the SHE
\cite{Sinova2004} in 2D electron systems have focused on linear
intrinsic Rashba- and/or Dresselhaus spin-orbit interaction (SOI) and
interpreted this boundary spin accumulation in terms of a spin current
\cite{Dyakonov1971} (defined as a symmetrized product of spin and
current densities) flowing transverse to the applied electric
field. However, these arguments have been plagued by ambiguities, such
as equilibrium spin currents \cite{Rashba2003b} and the absence of
spin conservation \cite{Erlingsson2005,Chalaev2005} in systems with
intrinsic SOI. Explicit diagrammatic calculations
\cite{Inoue2004,Raimondi2005,Chalaev2005} for disordered systems and a
more general, non-perturbative argument
\cite{Erlingsson2005,Chalaev2005, Dimitrova2005} show that the
spin-current is absent in systems with standard linear-in-momentum
SOI.\footnote{In quantum wells with more than one subband the spin-Hall current
can be nonzero. See Ref.\onlinecite{Lee2009}.} 

A more straightforward approach is to calculate the quantity directly
measured in experiments: the spatially and time resolved spin density.
\cite{Adagideli2005,Malshukov2005,Galitski2006} In weakly disordered
systems with $E_F \gg \tau^{-1}, \Delta_{\mathrm{SO}}$ (where $E_F$ is
the Fermi energy, $\tau$ the momentum relaxation time, and
$\Delta_{\mathrm{SO}}$ the spin-orbit splitting) the spin density is
described by spin diffusion equations derived in Keldysh
\cite{Mishchenko2004,Bernevig2006,Tserkovnyak2007} or density matrix
approaches.\cite{Burkov2004,Galitski2006,Adagideli2007} These
equations have been used to study various effects, such as the
response to an electromagnetic wave, \cite{Shnirman2007} spin
currents, \cite{Mishchenko2004} spin relaxation,
\cite{Pershin2005,Stanescu2007} boundary spin accumulation for
dc\cite{Adagideli2005}$^,$\cite{Malshukov2005,Rashba2006,Galitski2006,Bleibaum2006}
and abruptly switched\cite{Raimondi2007,Stern2008} electric fields,
and more general interface
problems. \cite{Tserkovnyak2007,Adagideli2007}

A significant difference between charge and spin diffusion, as
described by these equations, is the existence of spatially
oscillating spin density modes. For instance, a gradient of the
out-of-plane spin density acts as a torque on the in-plane spin and
vice versa, leading to a periodic spatial modulation of both in- and
out-of-plane spin densities with a period given by the spin-orbit
length $\lambda_{\mathrm{SO}}$. General solutions of the spin
diffusion equations are damped spatial spin density oscillations with
a period given by the spin-orbit relaxation length
$\lambda_{\mathrm{SO}}$.  An example of such periodic modes in
diffusive systems was first described in
Ref.~\onlinecite{Schliemann2003} (see, in particular, Eq.~(7) there)
for the case of equal strengths of the Rashba and linear Dresselhaus
SOI. For this particular case and in the absence of the cubic SOI,
these modes are long-lived and static and are thus referred to as
persistent spin helix.\cite{Bernevig2006} Modes of this type have
recently been observed.\cite{Koralek2009}

However, when analyzing these equations for a specific geometry, e.g.,
in a narrow channel for the case of the SHE, the weight of these
oscillatory modes in the solution is determined by boundary conditions
(BCs). For instance, assuming vanishing polarization at the boundary
one obtains an oscillatory behavior of the spin density,
\cite{Rashba2006} resembling the spin profile measured in
Ref.~\onlinecite{Sih2005}. On the other hand, for a von-Neumann
boundary condition (vanishing normal gradient of the polarization),
the spin profile is spatially uniform. Thus, the existence of the SHE
depends crucially on the BCs.  This circumstance motivated a number of
studies where BCs for systems with SOI were derived microscopically,
both in the diffusive \cite{Adagideli2005,Galitski2006, Bleibaum2006,
  Tserkovnyak2007, Rashba2006} and ballistic \cite{Zyuzin2007}
regimes.

It has been shown\cite{Bleibaum2006,Tserkovnyak2007} that BCs (for
hard-wall spin-conserving boundaries) in
disordered\footnote{Qualitatively different spin effects occur in
  systems with orbital phase coherence, i.e., in mesoscopic
  disordered\cite{Duckheim2008} or ballistic
  systems\cite{Bardarson2007,Adagideli2009} and on length scales
  smaller than the mean free path at ballistic
  boundaries.\cite{Silvestrov2009}} systems with linear SOI and for dc
electric fields require the spin density to be equal to its value in
the bulk, i.e., far away from the boundary, and, thus, lead to a
spatially uniform spin profile. This null result is consistent with
zero spin currents.  \cite{Inoue2004,Raimondi2005,Chalaev2005,
  Dimitrova2005,Mishchenko2004} The experimentally observed dc spin
accumulation \cite{Sih2005} in 2DEGs thus requires an explanation
accounting for both extrinsic \cite{Engel2005,Raimondi2009} and (cubic
\cite{Malshukov2005}) intrinsic effects.  That a spin current is
finite at finite frequencies and for linear SOIs,\cite{Mishchenko2004,
  Duckheim2006, Duckheim2007} however, hints at the presence of
boundary spin accumulation in ac solutions.  In this article, we focus
on the intrinsic mechanism, and show that a {\it dynamic} SHE, i.e.,
boundary spin polarization induced by an ac voltage, is present even
in a minimal intrinsic model.

The dynamic SHE arises due to the excitation of spatially non-uniform
spin diffusion modes. In the Hall-bar geometry, these modes are
excited by a spatially uniform ac electric field and lead to
accumulation and spatial oscillations of the spin density close to the
boundaries.  Analyzing these modes as a function of SOI strengths and
in the presence of an external, in-plane magnetic field, we find a
spin diffusion mode which is a finite-frequency analog of the
persistent spin
helix.\cite{Schliemann2003,Bernevig2006,Stanescu2007,Koralek2009} The
relaxation length of this mode -while finite for generic linear SOIs-
becomes infinite when the Rashba and Dresselhaus SOI strengths are
equal and when the magnetic field is tuned to resonance with the
frequency of the electric field.  This particularly robust mode,
originating from electric-dipole-induced spin resonance
(EDSR),\cite{Bell1962,Rashba2003}$^,$ \cite{Rashba1991, Kato2004,
  Schulte2005, Wilamowski2007,Meier2007,Duckheim2006, Duckheim2007}
gives rise to a spatially oscillating spin profile which extends
infinitely far away from the Hall-bar boundary. This driven spin helix
has the same spatial oscillation period as the persistent spin helix
\cite{Schliemann2003,Bernevig2006,Koralek2009} but, whereas the latter
is static, the former oscillates in time at the frequency of the
applied bias.  The prediction of a driven spin helix is one of the
main results of this paper.

Using a linear response approach, we solve the problem of a hard-wall
boundary in a disordered 2D electron gas in the presence of an ac
electric field. The derivation of the BCs is similar to the one in
Refs.~\onlinecite{Galitski2006} and \onlinecite{Bleibaum2006}. We find
that while the bulk polarization is reduced at finite frequencies, the
BCs require the polarization at the boundary to have a larger
value. The spin polarization is, thus, no longer spatially uniform:
there is a spin accumulation at the boundary and spatial oscillations
decaying towards the bulk of the sample. The amplitude of this spatial
oscillations at frequency $\omega$ is proportional to $\omega /\Gamma
$, where $\Gamma$ 
is the Dyakonov-Perel spin relaxation rate. Since typically $\Gamma
\ll \tau^{-1}$, where $\tau $ is the transport time, the dynamic SHE
becomes pronounced even for frequencies $\omega \tau \ll 1$.
Analyzing different geometries and SOIs, we find that it is possible
to excite a predominantly oscillatory mode for equal strength of the
Rashba and Dresselhaus SOI --a driven spin helix described above.

The paper is organized as follows. In Sec.~\ref{sec:model}, we
introduce our model and formulate the linear response formalism for
SHE.  In Sec.~\ref{sec:diffusion-equation}, we sketch the derivation
of the integral equation for the spin density, which is then used to
derive the diffusion equation and its boundary conditions.  (A more
detailed derivation is deferred to
appendix~\ref{sec:spin-diff-equat}.)  In
Sec.~\ref{sec:boundary-conditions-1}, we derive boundary conditions in
the presence of ac electric field and comment on the relation between
spin currents and these boundary conditions in
Sec.~\ref{sec:boundary-conditions}.  In
Sec.~\ref{sec:solut-diff-equat} we calculate the spatially resolved
spin profiles at finite frequencies in various geometries in $[001]$-
and $[110]$-grown quantum wells.  Generation of a driven spin helix
under the conditions of EDSR is discussed in
Sec.~\ref{sec:edsr-driven-spin} .

\begin{figure}
  \centering
  \includegraphics[width = 0.4 \textwidth]{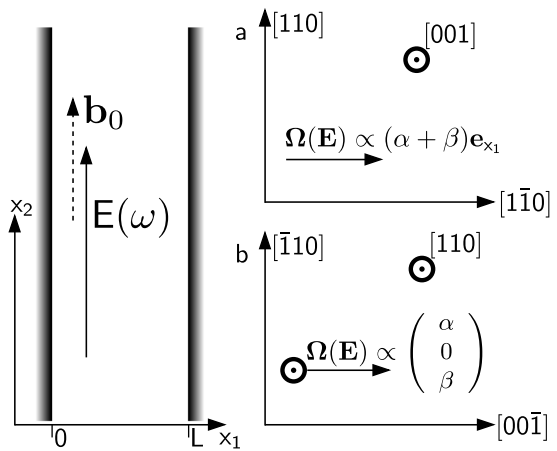}
  \caption{  Left: Conducting channel 
    infinite in the    $\mathbf e_{x_2}$-direction and of width $L$
    in the $\mathbf e_{x_1}$ direction. 
    ac Electric field $\mathbf E(\omega) || \mathbf e_{x_2}$ induces
    boundary spin accumulation.  An external magnetic field $\mathbf
    b_0$, applied parallel to $\mathbf E$, gives rise to EDSR (see
    Sec.~\ref{sec:edsr-driven-spin}).  Right, (a): a \lq\lq
    standard\rq\rq\/ $[001]$-grown quantum well with the $[110]$
    crystal axis taken along the $x_2$-direction. The bulk
    polarization $\boldsymbol{\Omega}(\mathbf e \mathbf E \tau)
    \propto \mathbf e_{x_1} (\alpha + \beta)$ points along $\mathbf
    e_{x_1}$ [cf. Eqs.~(\ref{eq:bulk-pol}) and (\ref{eq:SOI-001-gw})].
    Right, (b): a $[110]$-grown quantum well with $\mathbf E || [\bar
    110]$ along $\mathbf e_{x_2}$.  The internal field [see
    Eq.~(\ref{eq:SOI-110-gw})] $\boldsymbol{\Omega}(\mathbf e \mathbf
    E \tau)$ has both in-plane (due to the Rashba SOI) and
    out-of-plane (due to the Dresselhaus SOI) components.  }
  \label{fig:sample}
\end{figure}

\section{Preliminaries}\label{sec:model}

We consider a disordered 2DEG confined to a quantum well (QW) channel
of width $L$ (see Fig.~\ref{fig:sample}) with non-interacting
electrons of mass $m$ and charge $e$. The system is described by the
Hamiltonian
\begin{align}
\label{eq:hamiltonian}
H = \frac{\mathbf p^2}{2 m } + \boldsymbol{\Omega} (\mathbf p) \cdot
\boldsymbol{\sigma} + \mathbf b_0 \cdot \boldsymbol{\sigma} + V.
\end{align}
Here, $\mathbf p = (p_1, p_2,0 )$ is the in-plane momentum,
$\boldsymbol{\Omega}(\mathbf p)_i = \Omega_{ij} p_j$ is a linear,
vector-valued function of $\mathbf p$ describing spin-orbit
interaction, $2 \mathbf b_0 = g \mu_B (B_1,B_2,0)$ is a magnetic field
(equal in magnitude to the Zeeman energy) applied parallel to the
2DEG, and $\boldsymbol{\sigma} = (\sigma^1, \sigma^2, \sigma^3)$ are
the Pauli matrices (and $\sigma^0 = \mathbb{1}$). The disorder
potential $V$ due to static short-ranged impurities randomly
distributed over the channel is characterized by the mean free path $l
= \tau p_F/m $, where $\tau$ is the scattering time and $p_F$ is the
Fermi momentum.

We calculate the impurity-averaged, spatially-dependent spin density
$\hat{\scmd{S}}^i(\mathbf r) = \sigma^i \delta (\mathbf r -
\hat{\mathbf{x}})$ due to 
in-plane ac electric field $\mathbf E(\omega) = \mathbf E_0
\delta(\omega - \omega_0)$. As it will be shown below, the overall
magnitude of $\boldsymbol{\scmd{S}}$ is determined by the bulk spin
polarization due to CISP far away from the boundary. We therefore
briefly discuss CISP in different geometries. We define the nominal
polarization
\begin{align}
\label{eq:bulk-pol}
\mathbf{S}_{b} \equiv - \nu 2 \boldsymbol{\Omega}
(e\mathbf{E}(\omega)\tau) \, ,
\end{align}
(with $\nu = m/2 \pi$ being the density of states per spin) which at
zero frequency ($\omega_0 = 0$) coincides with the bulk
polarization.\cite{Edelstein1990} In this case, $\mathbf{S}_{b}$ is
simply a paramagnetic spin response to an effective magnetic field
$\boldsymbol{\Omega}(e \mathbf{E}_0 \tau)$. The latter is the internal
field due to the electrically induced drift momentum $e \mathbf E
\tau$ and SOI.

Both the magnitude and direction of $\mathbf S_b$ depend on the SOI
mechanism. We consider two cases (see Fig.~\ref{fig:sample}): the
``standard'' $[001]$- and $[110]$-grown QW. The Rashba SOI (with
strength $\alpha$) due to an asymmetry in the confinement potential
has the same form in both cases and is assumed to be tunable. The
Dresselhaus induced fields $\boldsymbol{\Omega_{D,[001]}}$,
$\boldsymbol{\Omega_{D,[110]}}$ are in-plane and out-of-plane in the
$[001]$ and $[110]$-grown QWs, respectively.  For convenience, we
define $\xi_{\alpha} = 2 \alpha p_F \tau$, $\xi_{\beta} = 2 \beta p_F
\tau$ as the ratios of the mean free path and spin precession length
due the Rashba and Dresselhaus SOIs, respectively.  The vector
couplings of the SOIs are described by
\begin{align}
  \label{eq:SOI-001-gw}
\Omega_{[001]} = \left(
  \begin{array}{ccc}
    0 & \alpha + \beta & 0 \\
    -(\alpha - \beta) & 0 &0 \\
    0 & 0 & 0 \\
  \end{array}
\right)
\end{align}
for case (a) in Fig.~\ref{fig:sample} and
\begin{align}
  \label{eq:SOI-110-gw}
\Omega_{[110]} =  \left(
    \begin{array}{lll}
      0 & \alpha  & 0 \\
      -\alpha  & 0 & 0 \\
      0 & \beta  & 0
    \end{array}
  \right)
\end{align}
for case (b).  In case (a), the bulk polarization $S_b \propto -
\mathbf e_{x_1} (\alpha + \beta)$ points along the (negative) $x_{1}
$-axis. When the Rashba- and Dresselhaus SOIs are of comparable
strength, i.e., $\alpha \approx +\beta$ (or $\alpha \approx - \beta$),
constructive (destructive) interference between the two SOI mechanisms
occurs.\cite{Schliemann2003} In this case, one spin component [along
$x_1$ ($x_2$)] becomes conserved. A similar situation occurs in the
$[110]$-grown QW, where the out-of-plane spin is conserved if the
Rashba SOI is relatively small. Here the bulk polarization points
out-of-plane and is, thus, easier accessible in optical measurements.
\cite{Kato2004b,Sih2005,Meier2007}

The induced spin density $\scmd{S}^\mu(\mathbf r)$ is described by
coupled spin diffusion equations
\cite{Mishchenko2004,Burkov2004,Malshukov2005} which can be derived in
the Keldysh \cite{Mishchenko2004,Bernevig2006,Tserkovnyak2007} or
density matrix
formalisms. \cite{Burkov2004,Galitski2006,Adagideli2007} As a starting
point for the derivation of the boundary conditions, we present here
an alternative derivation based on a diagrammatic linear response
approach. The detailled derivation is deferred to
Appendix~\ref{sec:spin-diff-equat}. We obtain an integral equation for
the spin density
\begin{align}
\label{eq:rho-integral-eq3}
  \scmd{S}^{i}(\mathbf{r}) - S_{b}^{i} & = i \omega \tau
  S_{b}^{i} +   \int d^{2}x\,
  X^{i j}(\mathbf{r},\mathbf{x}) (\scmd{S}^{j}(\mathbf{x}) - S_{b}^{j}) \, 
\end{align}
valid in the regime $1/E_F \tau \ll 1$, where
\begin{align}
  \label{eq:spin-spin-diagram}
  X^{\mu \nu}(\mathbf{r},
  \mathbf{x})=\frac{1}{2m\tau}\mathrm{tr}\left\{
    \sigma^{\mu}G^{R}_{E_F+\omega}(\mathbf{r},\mathbf{x})\sigma^{\nu}
    G^{A}_{E_F}(\mathbf{x}, \mathbf{r}) \right\} .
\end{align}
Here, $E_F$ is the Fermi energy, $\mathrm{tr}\{ \dots \}$ denotes the
trace over spin s, $ \hat{v}_{j}=\frac{\hat
  p_{j}}{m}+\Omega_{kj}\sigma^{k}$ is the velocity operator containing
a spin-dependent term, and $G^{R/A}_E$ the impurity-averaged,
retarded/advanced Green functions at energy $E$. Note that for
$\omega=0$ the integral equation~(\ref{eq:rho-integral-eq3}) depends
only on the combination $\boldsymbol{\scmd{S}} -\mathbf S_b$ so that
the spatially uniform solution $\boldsymbol{\scmd{S}} = \mathbf S_b$
is immediate. The uniform spin profile is equivalent to the absence of
the SHE, whose presence would cause a spatial modulation of $\scmd{S}$
at the boundary.

\begin{figure}
  \centering
  \includegraphics[width = 0.48 \textwidth]{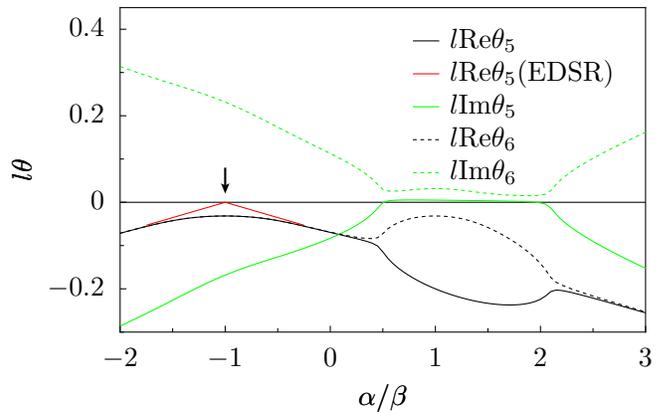}
  \caption{(color online) Characteristic wave numbers $\theta$ of the
    homogeneous solutions $s_5(r_1)= s_{5,0} e^{\theta_5 r_1}$ and
    $s_6=s_6(r_1)= s_{6,0} e^{\theta_6 r_1}$ of Eq.~(\ref{eq:rho-pde})
    as a function of $\alpha/\beta$ for $\omega \tau = 10^{-3}$ and
    $\xi_\beta = 0.1$ in a $[001]$-grown quantum well. For $\alpha = -
    \beta$ (indicated by arrow), the wave numbers have small real
    parts $| \mathrm{Re} \theta_5 |\approx | \mathrm{Re} \sqrt{-2 i
      \omega \tau}| \ll 1$, which implies a nearly undamped
    oscillatory mode. Under EDSR conditions, $\mathrm{Re} \theta_5$
    vanishes identically at $\alpha + \beta = 0$
    [cf. Eq.~(\ref{eq:theta-edsr})]. The presence of modes with almost
    imaginary wave numbers leads to an oscillating spin profile, as
    shown in Fig.~\ref{fig:polarization-plot001}.}
  \label{fig:mode}
\end{figure}

\section{Diffusion equation}\label{sec:diffusion-equation}

Far from the sample boundary, the impurity-averaged Green functions
and, hence, the kernel $X^{\mu\nu}(\mathbf r, \mathbf x') \approx
e^{-|\mathbf r - \mathbf x'|/l}$ in Eq.~(\ref{eq:rho-integral-eq3})
decay on the scale of the mean free path $l$, which is the shortest
length scale of the diffusion problem. The behavior of $\scmd{S}$ on
scales larger than $l$ can, thus, be found by expanding:
$\boldsymbol{\scmd{S}}(\mathbf{x}) \approx
\boldsymbol{\scmd{S}}(\mathbf{r})+ (\mathbf{x}-\mathbf{r})_{i}
\partial_{r_{i}} \boldsymbol{\scmd{S}}(\mathbf{r})+\frac{1}{2}
(\mathbf{x}-\mathbf{r})_{k}(\mathbf{x}-\mathbf{r})_{l}
\partial_{r_{k}}\partial_{r_{l}} \boldsymbol{\scmd{S}}(\mathbf{r})$.
In this way, one obtains the coupled spin diffusion equation
\begin{align}
  \label{eq:rho-pde}
  &[-i\omega + \Gamma -D\Delta_{\mathbf
    r} ] (\boldsymbol{\scmd{S}} (\mathbf{r})-
  \mathbf{S}_{b}) \\ \notag &- 2[\mathbf{b} -  p_{F}
  \boldsymbol{\Omega}(l\nabla_{r})] \times
  (\boldsymbol{\scmd{S}}(\mathbf{r})-\mathbf{S}_{b}) =i \omega \mathbf S_b
  \, ,
\end{align}
where $D=v_{F}l/2$ is the diffusion constant and $\Gamma^{ij}=
[\mathrm{tr}\{(\Omega\Omega^{T})\}\delta_{ij} -
(\Omega\Omega^{T})_{ij}]2p_{F}^{2} \tau$ is the spin relaxation
tensor.

We now apply Eq.~(\ref{eq:rho-pde}) to the two specific geometries in
Fig.~\ref{fig:sample} (a) and (b). Assuming translational invariance
along $\mathbf e_{x_2}$, we find for the $[001]$-grown QW with
$\mathbf E|| [110] || \mathbf e_{x_2}$
\begin{align}
\label{eq:rho-pde-001-1}
\left[ - i \omega -D \partial_{r_1}^2    + \Gamma_-
\right]  (\scmd{S}^1 -
S_b) + C_- \partial_{r_1}  \scmd{S}^3 &= i \omega S_b \, ,\\
\label{eq:rho-pde-001-2}
\left[ - i \omega - D \partial_{r_1}^2 + \Gamma_+
\right] 
\scmd{S}^2 &= 0 \, ,\\
\label{eq:rho-pde-001-3}
\left[ - i \omega - D \partial_{r_1}^2 + \Gamma_+
  + \Gamma_- \right] 
\scmd{S}^3 - C_- \partial_{r_1} \scmd{S}^1 &= 0  \, ,
  \end{align}  
where $\Gamma_{\pm} = 2 p_F^2 \tau (\alpha \pm \beta)^2$, $C_{\pm} =
2 p_F l (\beta \pm \alpha)$, $\omega_L = 2 b_0$ and $S_b = -2 \nu e E \tau (\alpha + \beta)$.

In case (b) of a $[110]$-grown QW with $\mathbf b_0, \mathbf E|| [\bar
110] || \mathbf e_{x_2}$ we find
\begin{align}
  \label{eq:rho-pde-110}
  [-D\partial_{r_{1}}^{2}-i\omega+ \Gamma'_{1}+
  \Gamma'_{2}] & (\scmd{S}^{1}-S_{b}^{1}) \notag  \\ - [C'_{2}\partial_{r_{1}}+
  \omega_{L}+ \sqrt{\Gamma'_{1}\Gamma'_{2}}] & (\scmd{S}^{3}-S_{b}^{3}) = i
  \omega S_{b}^{1} \, , \\
[-D\partial_{r_{1}}^{2}-i\omega+ \Gamma'_{1}+\Gamma'_{2}] & \scmd{S}^{2} =
0 \, , \\
[-D\partial_{r_{1}}^{2}-i\omega+  2 \Gamma'_{2}] & (\scmd{S}^{3}-S_{b}^{3})
\notag \\ +[C'_{2}\partial_{r_{1}}+\omega_{L}-\sqrt{\Gamma_{1}\Gamma_{2}}] &
(\scmd{S}^{1}-S_{b}^{1}) =
i\omega S_{b}^{3} \, ,
\end{align}
where $\mathbf{S}_{b}=- 2 \nu e E \tau(\alpha,0,\beta)$, $\Gamma'_{1}=
2 p_{F}^{2} \tau\beta^{2}$, $\Gamma'_{2}=2p_{F}^{2} \tau \alpha^{2}$,
$C'_{1}=p_{F}l\beta$, $C'_{2}=p_{F}l\alpha$, and $\omega_L= 2 b_0$.
Note that in this geometry the Dresselhaus SOI adds to $\mathbf
S_{b}$, whereas for $\mathbf E|| [001]$ the electric field does not
couple to the Dresselhaus term.\cite{Winkler2003}

\begin{figure}
  \centering
  \includegraphics[width = 0.48 \textwidth]{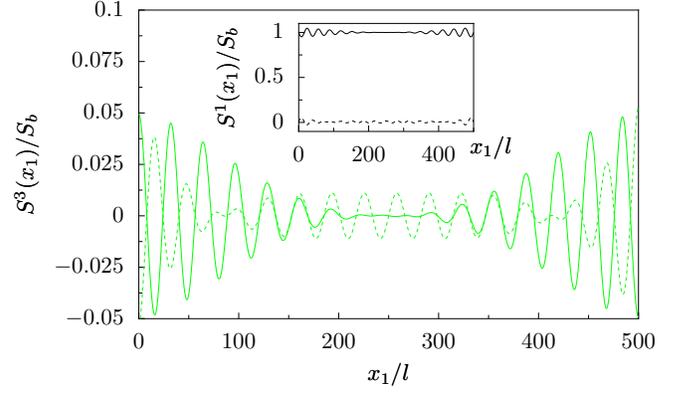}
  \caption{(color online) Real (solid line) and imaginary parts
    (dashed line) of the out-of-plane spin density $\scmd{S}^3(x)/S_b$
    [solution of
    Eqs.~(\ref{eq:rho-pde-001-1})-(\ref{eq:rho-pde-001-3}),~(\ref{eq:BC-explicit})]
    in the standard QW (Fig.~\ref{fig:sample} a) are shown for $\omega
    \tau= 10^{-4}$, $\xi_\alpha =0.1$, $\xi_\beta =- 0.095$, and $L=
    500 l$.  As Rashba- and Dresselhaus SOI interfere destructively
    for this case, the bulk polarization is much smaller than the
    value discussed in Sec.~\ref{sec:solut-diff-equat}: all other
    parameters being equal, $S_b = 0.005\; \mu \mathrm m^{-2}$ instead
    of $1.1\; \mu \mathrm m^{-2}$. Inset: In-plane polarization
    $\scmd{S}^1(x)$ (along the internal field $\boldsymbol{\Omega}(e
    \mathbf{E}\tau)$) in the same situation. }
  \label{fig:polarization-plot001}
\end{figure}

\section{Boundary conditions}\label{sec:boundary-conditions-1}

The diffusion equation~(\ref{eq:rho-pde}) has to be supplemented with
boundary conditions. These match the bulk solutions of the diffusion
equation~(\ref{eq:rho-pde}) with the solution of the integral equation
Eq.~(\ref{eq:rho-integral-eq3}) in the region $1/p_F\ll x_1 \ll l$
close to the boundary. Here, we follow the approach used in
Refs.~\onlinecite{Galitski2006} and \onlinecite{Bleibaum2006}. We
choose $x_1=0$ as the boundary and construct the impurity-averaged
Green functions $G^{R/A}(\mathbf x, \mathbf x')$ which satisfy the
Dyson equation $\langle \mathbf x |[(\omega - \hat{H}_0-\hat{H}_{SO} -
\hat{\Sigma}) \hat{G}] | \mathbf x' \rangle =\delta(\mathbf x -
\mathbf x')$ with $H_0$ being the Hamiltonian in the absence of SOI
and $\hat{\Sigma}$ being the self-energy due to impurity
scattering.\cite{Akkermans2007,Rammer1986} We, moreover, impose the
hard-wall, spin-conserving boundary conditions $G(\mathbf x, \mathbf
x')|_{x_1,x'_1=0} = 0$ for either argument at the boundary.

To 0th order in the SOI, these conditions are satisfied by image
constructions
$G^{R/A}_0=G_{\mathrm{b},0}^{R/A}-G_{\mathrm{b},0}^{\ast R/A}$,
where $G^{ R/A}_{\mathrm{b,0}}$ is the impurity-averaged Green
function in the bulk and $G^{\ast R/A}_{\mathrm{b}}(\mathbf x,
\mathbf x') = G_{\mathrm{b}}^{R/A}(\mathbf{x}, (-x'_{1},x'_{2}))$ is
the Green function mirror-reflected at the boundary. Neglecting
Friedel oscillations of the self-energy at the boundary, which fall
off as $1/ \sqrt{p_F x_1}$, the Green functions $G^{R/A}$ constructed
in this way satisfy the Dyson equation to leading order in $1/E_F
\tau$.

To 1st order in $H_{SO}$ the Green functions is found as $\hat{G}_1 =
\hat{G}_0 \hat{H}_{SO} \hat{G}_0$. By construction, $G(\mathbf x,
\mathbf x') = [G_0 + G_1](\mathbf x, \mathbf x')$ satisfies the
boundary conditions and the Schroedinger equation to linear order in
the spin-orbit interaction. Performing a Fourier transform of the
Green function $G(\mathbf x, \mathbf x') = \int dp_2 G(x_1,x_1'|p_2)
e^{i p_2 (x_2 - x_2')} /(2 \pi)$ along the boundary, we find
\begin{align}
  \label{eq:green-functions-0}
G_0^{R/A} (x_1,x_1'|p_2) = \frac{\mp i m}{p_E^{\pm}} \left[ e^{\pm i p_E^{\pm}
  |x_1 - x_1'| }  - e^{\pm i p_E^{\pm}
  (x_1 + x_1') } \right], 
\end{align}
where $p_E^{\pm} = \sqrt{2 m (E \pm i/2 \tau - p_2^2 / 2m)}$ with
$p_2$ being the momentum along the channel. To first order in
$H_{\mathrm{SO}}$, we find
\begin{align}
  \label{eq:green-functions-1}
G_1^{R/A} (x_{1},x_{1}'|p_2) &= \frac{\mp m^2 \Omega_{k1} \sigma^k}{p_E^{\pm}}
(x_{1}-x_{1}') \\ &\times  \left[e^{\pm i p_E^{\pm}  |x_{1} - x_{1}'|} - e^{\pm i p_E^{\pm}
  (x_{1} + x_{1}')} \right] + \dots \notag , 
\end{align}
where the dots stand for additional terms that do not contribute to
the integrals below since they are odd in the longitudinal momentum
$p_2$.

We are now in a position to derive boundary conditions using the Green
functions from Eqs.~(\ref{eq:green-functions-0}) and
(\ref{eq:green-functions-1}).  We take the limit $\mathbf{r} \to
0$ of Eq.~(\ref{eq:rho-integral-eq3}) and expand
$\boldsymbol{\scmd{S}}^{i}(\mathbf{x}) \approx
\boldsymbol{\scmd{S}}^{i}(\mathbf{r}) + (x_{j}-r_{j})
\frac{\partial}{\partial r_{j}} \boldsymbol{\scmd{S}}^{i}(\mathbf r)$
in the integrand. This yields
\begin{align}
  \label{eq:boundary-cond}
  0&= i\omega \tau \mathbf{S}_{b}  + (
  B - \mathbb 1) (\boldsymbol{\scmd{S}} ( 0) - \mathbf{S}_{b} )
  + C_{j}\frac{\partial}{\partial
    r_{j}} \boldsymbol{\scmd{S}}(0) ,
\end{align}
where the coefficients
\begin{align}
  \label{eq:boundary-coefficients}
  B^{\mu\nu} &= 
  \int_{x'_{1}>0}dx'_{1} dx'_{2} X^{\mu\nu}(\mathbf{x},
  \mathbf{x'})\Large|_{x \to 0} \\ C^{\mu\nu}_j &=
  \int_{x'_{1}>0}dx'_{1}dx'_{2}X^{\mu\nu}(\mathbf{x},\mathbf{x'})(x'_{j}-x_{j})\Large|_{x\to 0} 
\end{align}
are obtained from the spin-spin correlation function $X^{\mu\nu}$ in
Eq.~(\ref{eq:spin-spin-diagram}) evaluated with the Green's functions
satisfying the boundary conditions.  In symbolic notations, $X \propto
G_{\mathrm{b},0}^RG_{\mathrm{b},0}^A + G_{\mathrm{b},0}^{\ast R}
G_{\mathrm{b},0}^{\ast A} - G_{\mathrm{b},0}^R G_{\mathrm{b},0} ^A-
G_{\mathrm{b},0}^{\ast R} G^A_{\mathrm{b},0}$.  Note that for
$\omega=0$ the diffusion equation Eq.~(\ref{eq:rho-pde}) and the
boundary conditions Eq.~(\ref{eq:boundary-cond}) depend only on the
combination $\boldsymbol{\scmd{S}} -\mathbf S_b$, so that the
spatially constant solution $\boldsymbol{\scmd{S}} = \mathbf S_b$ is
immediate.  In particular, there is no spin accumulation close to the
boundary in that case in agreement with the literature on the linear
intrinsic spin-Hall effect. \cite{Inoue2004, Raimondi2005,
  Malshukov2005, Chalaev2005}

When calculating the spin-spin correlation function
$X^{\mu\nu}$
one encounters mixed terms of the form $G G^{\ast}$, which oscillate
as a function of $x_1$ with a period of $1/p_F$. To determine
$\scmd{S}$ on length scales larger than $l$, we neglect these
oscillations. This way, we find the BCs
\begin{align}
  \label{eq:BC-explicit}
l \partial_{\mathbf{\hat n } } \scmd{S}^i = - 2 p_F \tau
\Omega(\mathbf{\hat n } )_m \epsilon_{m i j} \left(\scmd{S}^j - \scmd{S}_b^j\right)  \, ,
\end{align}
where $\mathbf{\hat{ n}}$ is a unit vector normal to the boundary and
where we have neglected terms proportional to $\omega\tau\ll1$.

\section{Spin current}\label{sec:boundary-conditions}

In this section, we show that a definition of the spin current in
terms of an SU(2)-covariant derivative is consistent with both the
boundary conditions and the diffusion equation. This definition is
equivalent to defining the spin current as the commutator -- in contrast
to the conventionally used \emph{anti}commutator -- of spin and velocity. To see
this, we define a Hermitian spin current operator as follows
\begin{align}
\label{eq:spin-current-comm-op}
\hat{J}_{i}^{\eta}( \boldsymbol{\hat{\scmd{S}}}) &=  - \mathcal{D}_i
\hat{\scmd{S}}^{\eta}(\mathbf{r})  = -i [m \hat v_{SO,i}, \hat{\scmd{S}}^{\eta}(\mathbf{r})] 
\\ \notag &=
\frac{\partial}{\partial r_{i}} \hat{\scmd{S}}^{\eta}(\mathbf{r}) + 2 m
\Omega_{ki}\epsilon_{k\eta\eta'} \hat{\scmd{S}}^{\eta'}(\mathbf{r}) ,
\end{align}
where we have introduced the covariant derivative\cite{Tokatly2008}
$\mathcal{D}_{i}\,\underline{\cdot}= \partial /\partial \hat x_{i} \,
\underline{\cdot}-i[\mathcal{A},\,\underline{\cdot}]$ with the
non-abelian gauge potential $\mathcal{A}_{i}=- m\Omega_{ki}\sigma^{k}$
and $\hat{\scmd{S}}^\eta(\mathbf r) = \sigma^\eta \delta (\mathbf r -
\hat{\mathbf{x}})$ is the spin density operator. [Note that
$\hat{J}_{i}^{\eta}$ differs by a factor of mass $m$ from the
conventionally defined product of velocity and spin.]
From Eq.~(\ref{eq:spin-current-comm-op}) we obtain a spin current $J$
by replacing $\boldsymbol{\hat{\scmd{S}}}$ by
$\boldsymbol{\scmd{S}}(\mathbf r) - \mathbf S_b$ in the second line of
Eq.~(\ref{eq:spin-current-comm-op}), i.e.,
\begin{align}
\label{eq:spin-current-comm}
J_{i}^{\eta}(\mathbf r) 
= -\mathcal{D}_{i}^{\eta \eta'}\left(S^{\eta'}(\mathbf r) - S_{b}^{\eta'}\right)
\, ,
\end{align}
where $-\mathcal{D}_{i}^{\eta \eta'}=\delta^{\eta \eta'} \partial /\partial r_{i}
+ 2m\Omega_{ki}\epsilon_{k \eta \eta'}$. The BCs in
Eq.~(\ref{eq:BC-explicit}) are then equivalent to the requirement that
the normal component of ${\mathbf J}$ vanishes at the boundary, i.e.,
$\mathbf{\hat n}\cdot\mathbf{J}^{\eta}(\mathbf r) \Large|_{r_1 \to
  0}=0 \, , \eta =1,2,3$.

Using the definition, Eq.~(\ref{eq:spin-current-comm}), one finds that
both the diffusion equation [for this see also
Ref.\onlinecite{Raimondi2009}], Eq.~(\ref{eq:rho-pde}), and the
boundary conditions, Eq.~(\ref{eq:BC-explicit}), can be written in
terms of the covariant derivative as
\begin{align}
  \label{eq:diff-eq-bc-covariant-der-1}
-i \omega S^{\eta} + D \mathcal{D}_{i}^{\eta
  \eta'} J_{i}^{\eta'} &= 0 \\ 
  \label{eq:diff-eq-bc-covariant-der-2}
\hat{\mathbf{n}}\cdot\mathbf{J}^{\eta} \Big|_{r_1 \to 0} & =0 \, .
\end{align}
Thus, spin diffusion with linear SOI has a (formal) analogy to charge
diffusion: In charge diffusion, both the diffusion equation $\dot
\rho=D \nabla \mathbf j$ for the charge density $\rho$ and the BCs
$\hat{\mathbf{n}} \cdot \mathbf j =0$ contain the same charge current
$\mathbf j$. The current $\mathbf j = \nabla \rho$ is given in terms
of the spatial derivative of the density. Analogously, the spin
current is given as the SU(2)-covariant derivative of $S^{\eta}$.

In Ref.~\onlinecite{Bleibaum2006}, in an attempt to identify a spin
current directly from the diffusion equation, Eq.~(\ref{eq:rho-pde})
was rewritten (for $\mathbf b_0=0$) in the form
\begin{align}
  \label{eq:de-with-spin-current}
  -i \omega \scmd{S}^\eta +[\Gamma
  (\scmd{S} - S_b)]^\eta - D \nabla \cdot \tilde{\mathbf{J}}^\eta = 0 \, ,
\end{align}
where the \lq\lq spin current\rq\rq\/
\begin{align}
  \label{eq:spin-current-de}
\tilde{J}^\eta_i = \frac{\partial}{\partial r_{i}}
\hat{\scmd{S}}^{\eta}(\mathbf{r})- 4 m \Omega_{ki} \epsilon_{k \eta'\eta} \hat{\scmd{S}}^{\eta'}(\mathbf{r}) ,
\end{align}
however, differs from $J^\eta_i$ by a relative factor of 2. This
discrepancy is resolved when the definitions
Eqs.~(\ref{eq:spin-current-comm-op}),~(\ref{eq:spin-current-comm}) are
used making the introduction of two different spin currents $J$ and
$\tilde J$ unnecessary.

\section{Solutions of the diffusion equation}\label{sec:solut-diff-equat}

\begin{figure}
  \centering
  \includegraphics[width = 0.48 \textwidth]{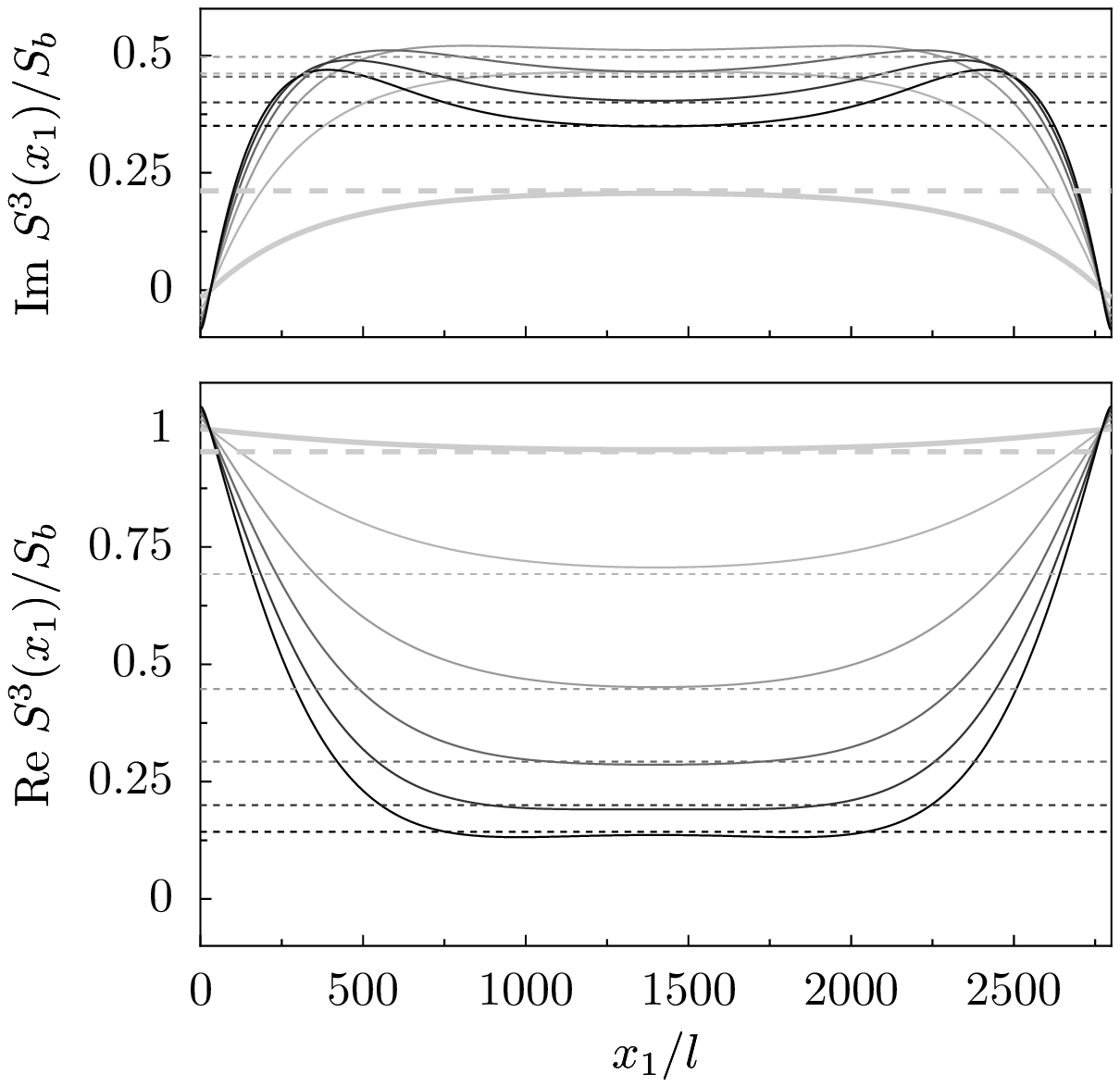}
  \caption{Real(upper panel) and imaginary parts (lower panel) of the
    out-of-plane spin polarization $\scmd{S}^3(x)/S_b$ for frequencies
    $\omega \tau= 0.1, 0.3, 0.5, 0.7, 0.9, 1.1 \times 10^{-5}$ (grey
    to black lines), $\xi_\beta =0.1$, $\xi_\alpha =0.003$, $L= 2800
    l$, and $S_b = 1 \mu \mathrm m^{-2}$.  The geometry with $\mathbf
    E || [\bar 110]$ is shown in Fig.~\ref{fig:sample} (b).
    Horizontal dashed lines mark ac bulk polarization according to
    Eq.~(\ref{eq:rho-in}) for the same parameters.}
  \label{fig:polarization-plot110}
\end{figure}

First, we obtain a solution of Eq.~(\ref{eq:rho-pde}) in an infinite
sample. In this case the bulk Green's functions $G_{\mathrm{b}}^{R/A}
(\mathbf{x},\mathbf{x}') =
G_{\mathrm{b}}^{R/A}(\mathbf{x}-\mathbf{x}')$ are translationally
invariant and, thus, $X^{\mu \nu} \equiv \int d^{2}x'X^{\mu
  \nu}(\mathbf{r,}\mathbf{x}')$ becomes independent of $\mathbf
r$. The spatially uniform ansatz
\begin{align}
  \label{eq:rho-in}
  \boldsymbol{\scmd{S}}_{\infty}&= 
  [\Gamma - i\omega ]^{-1} \Gamma \mathbf{S}_{b}
\end{align}
solves both the integral equation [Eq.~(\ref{eq:rho-integral-eq3}] and
the diffusion equation [Eq.~(\ref{eq:rho-pde})]. The same result for
the polarization at finite $\omega$ was found in
Ref.~\onlinecite{Raichev2007} using a kinetic equation and in
Refs.~\onlinecite{Duckheim2006, Duckheim2007} in the linear response
formalism.  Remarkably, $\boldsymbol{\scmd{S}}_{\mathrm{\infty}}$ is
not simply given by the ac internal field $\boldsymbol{\Omega}
(\mathbf p_d)$ corresponding to ac drift momentum $\mathbf p_d = e
\mathbf E (\omega) \tau / (1 - i \omega \tau)$, but depends on the
spin relaxation rate.  Therefore, the deviation of
$\boldsymbol{\scmd{S}}_{\infty}$ from $\mathbf S_b$ becomes
appreciable already at a relatively small frequency $\omega \simeq
\Gamma_{DP}$ rather than at a much higher frequency $\omega \simeq
\tau^{-1}$, which marks the dispersion of ${\mathbf p}_d$.  Note also
that there is no dc bulk polarization at $|\alpha| = |\beta|$, if the
limit of $\alpha \to \pm \beta$ is taken before the limit of $\omega
\to 0$ [see Ref.~\onlinecite{Duckheim2009} for a discussion of this
point].

We now estimate the magnitude of $\mathbf S_b$. We choose parameters
similar to the (low-mobility) sample employed in
Ref.~\onlinecite{Sih2005} except for a higher mobility and a lower
sheet density. With $\alpha = 1.0 \times 10^{-12}\; \mathrm{eVm}$,
sheet density $n_2 = 1.0 \times 10^{15}\; \mathrm{m}^{-2}$, and
transport mean free path $\tau = 5 \times 10^{-13} \;\mathrm s$ and
choosing $E = 5 \;\mathrm{m V}/\mu \mathrm m$, we obtain the bulk
polarization due to Rashba SOI $S_{b, \alpha} \equiv 2 \nu e E \tau
\alpha =1.1 \mu \mathrm m^{-2}$, or about 1 spin per $\mu \mathrm
m^{2}$ ($~S_b/n_2 = 0.1 \%$). The magnitude of $\scmd{S}_{\infty}$
and, as we will see below, the magnitude of the spatially non-uniform
terms in the solution are proportional to $S_b$.  Depending on the
geometry and on whether the Rashba and Dresselhaus SOIs add
constructively or destructively, the overall amplitude of the spin
oscillations and edge spin accumulation is modified.  In case (a) in
Fig.~\ref{fig:sample}, one finds $S_b = S_{b,\alpha} (1 +
\beta_{[001]}/ \alpha )$ while in case (b) $\mathbf S_b = S_{b,
  \alpha} (1,0, \beta_{[110]}/ \alpha)$, where $\beta_{[001]}$ and
$\beta_{[110]}$ is the Dresselhaus SOI strength in the [001]- and
[110]-grown QW, respectively.

We now focus on the position-dependent spin profile in a semiconductor
channel of finite width.  As before, we assume translational
invariance along the channel so that the diffusion equation
Eq.~(\ref{eq:rho-pde}) becomes an inhomogeneous ordinary differential
equation
\begin{align}
  \label{eq:DE-operator}
L(\partial_{r_1})
[\boldsymbol{\scmd{S}} (\mathbf{r}_1) - \mathbf{S}_{b}] =
i\omega\tau\mathbf{S}_{b}
\end{align}
in the transverse coordinate $r_1$, where the differential operator
$L(\partial_{r_1})$ is defined by Eqs.~(\ref{eq:rho-pde}) and (\ref{eq:DE-operator}).
The solution
\begin{align}
\label{eq:general-solution}
 \boldsymbol{\scmd{S}} = \boldsymbol{\scmd{S}}_{\infty} +
c^{k}\mathbf{s}_{k}(\mathbf{r}) 
\end{align}
consists of the uniform part $\boldsymbol{\scmd{S}}_{\infty}$, given by
Eq.~(\ref{eq:rho-in}) (inhomogeneous solution), and a linear
combination of $k=1,2,\dots,6$ eigenmodes $\mathbf{s}_{k} = \mathbf
s_{k,0} e^{\theta_k r_1}$, satisfying
$L(\nabla_{r})\mathbf{s}(\mathbf{r})=0$. The wave numbers
$\theta_{1,\dots,6}$ (in arbitrary order) in case (a)
are given by\footnote{Below,
  we focus on the behavior of $S$ for $\alpha \approx + \beta$
  ($\alpha \approx - \beta$). In a $[110]$-grown quantum well, similar
  behavior can be found when the Rashba SOI strength is close to zero
  and the electric field is taken along the $[\bar 110]$-direction
  (the $[001]$-direction). Therefore, we discuss case (a) in more detail.}
\begin{align}
  \label{eq:modes-theta}
  \theta_{1,2 } &= \mp l^{-1} \sqrt{2 \tau \left(\Gamma_+-i \omega
    \right)} \notag \\
  \theta_{3,4, (5,6)} &= \frac{+(-)1}{2 D} \Bigg[2 D \left(
    \Gamma_+ - 2 i
    \omega \right)  -C_-^2  \\ & +(-) \mp  2 \sqrt{-2 D
    \left(\Gamma_+ -2 i \omega \right) C_-^2+D^2 \Gamma_+^2}
  \Bigg]^{\frac{1}{2}}  \notag 
\end{align}
Some of $\theta_k$ are shown in Fig.~\ref{fig:mode} as functions of
$\alpha / \beta$. The real and imaginary part of the wave number
are responsible for exponentially growing (decaying) and oscillatory
parts of the mode, respectively. The coefficients $c^{k}$ are
determined by the boundary conditions in the form $M \mathbf c = -
\left( (B -\mathbb 1) (\boldsymbol{\scmd{S}}_{\infty} - \mathbf S_b),
  -(B -\mathbb 1) (\boldsymbol{\scmd{S}}_{\infty} - \mathbf S_b)
\right)$ where $M$ is a $6\times6$ -matrix obtained by inserting the
general solution into Eq.~(\ref{eq:boundary-cond}) [see also
Eq.~(\ref{eq:BC-matrix-form}) in
Appendix~\ref{sec:boundary-conditions-2}]. The coefficients $c^{k}$
determine the magnitude of the non-uniform part of $\scmd{S}$, i.e.,
if all $c^{k}$ are zero the solution is spatially uniform. Although
explicit expressions for $c$ are too lengthy to be displayed here, the
scaling of $c$ with $\omega$ can be found on general grounds. Indeed,
all the entries of the matrix $M^{-1} \mathrm{diag}((B -\mathbb 1),
-(B -\mathbb 1))$ are of order $1$.  The order of magnitude of $c_k$
is thus given by $| \boldsymbol{\scmd{S}}_{\infty} - \mathbf S_b|
\approx (\omega/\Gamma)|\mathbf S_b|$, where the latter holds for
$\omega \lesssim \Gamma$. The non-uniform part of $\scmd{S}$
(proportional to the $c$'s), thus, scales linearly with $\omega$ for
$\omega\ll\Gamma$ and becomes appreciable at the frequency scale $
\omega \simeq \Gamma \ll \tau^{-1} $.

A solution for $\boldsymbol{\scmd{S}}$ in a $[001]$-grown QW
(Fig.~\ref{fig:sample} a) is shown in
Fig.~\ref{fig:polarization-plot001}.  The electric field $\mathbf E$
is along the $[1 1 0]$ axis and the strengths of the Rashba and
Dresselhaus SOIs are chosen as $\alpha \approx - \beta$, so that the
wave numbers (cf.  Fig.~\ref{fig:mode}) are almost imaginary. In this
case, oscillations of the out-of-plane spin density $\scmd{S}^3$
extend almost over the entire channel. Simultaneously with $\alpha$
approaching $ - \beta$, however, the internal field
$\boldsymbol{\Omega}(e \mathbf E \tau) \propto \alpha + \beta$ and,
thus, the overall amplitude $S_b$ of the spin density becomes
small. In other words, suppression of the damping rate $\mathrm{Re}
\theta_i \propto |\alpha + \beta|$ close to the special point $\alpha
= -\beta$ competes with a suppression of the overall amplitude, so
that a purely oscillatory mode cannot be excited in this geometry.

Figure~\ref{fig:polarization-plot110} depicts the polarization profile
in a wide $[110]$-grown QW as shown in Fig.~\ref{fig:sample} (b),
where the bulk polarization (due to the Dresselhaus SOI) is
out-of-plane.  For a weak Rashba SOI, the wave numbers of the
characteristic modes are almost real, i.e., the modes are strongly
damped. As a result, the polarization close to the boundary is
substantially larger than the bulk value given by
Eq.~(\ref{eq:rho-in}).

\begin{figure}
  \centering
  \includegraphics[width = 0.45
  \textwidth]{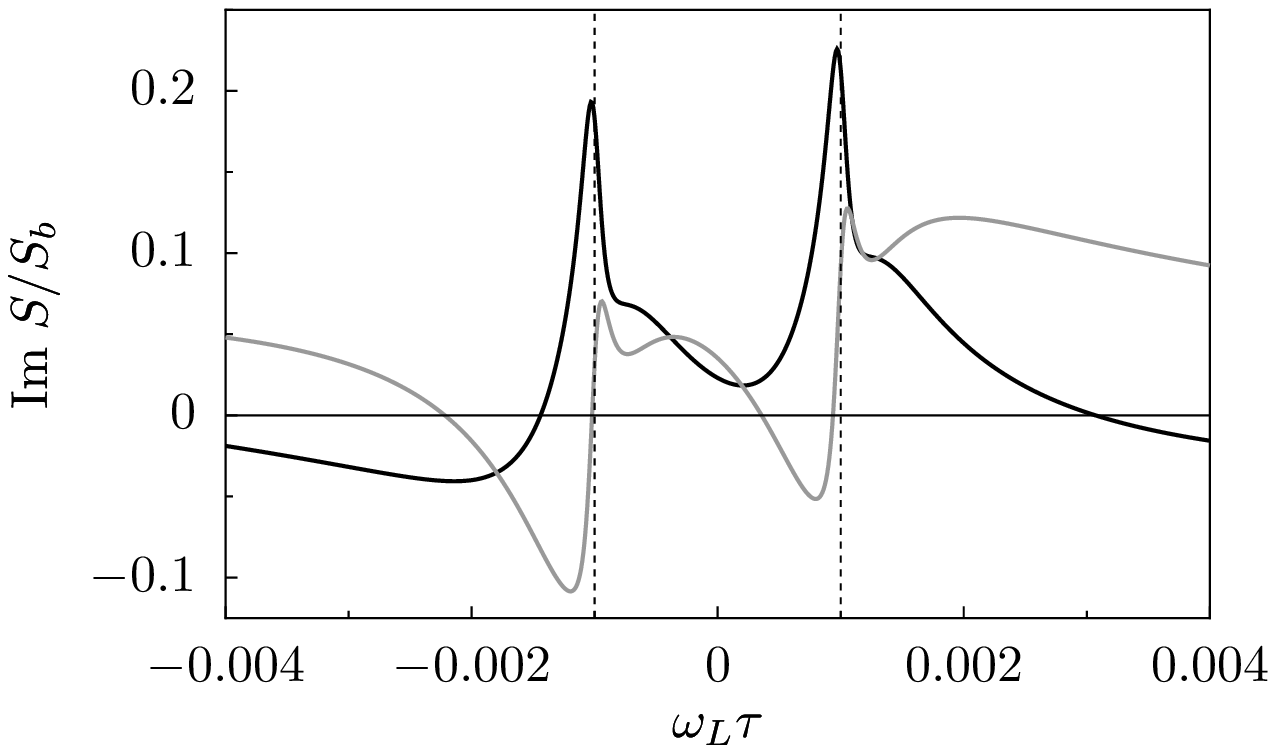}\\
  \bigskip
  \includegraphics[width = 0.22 \textwidth]{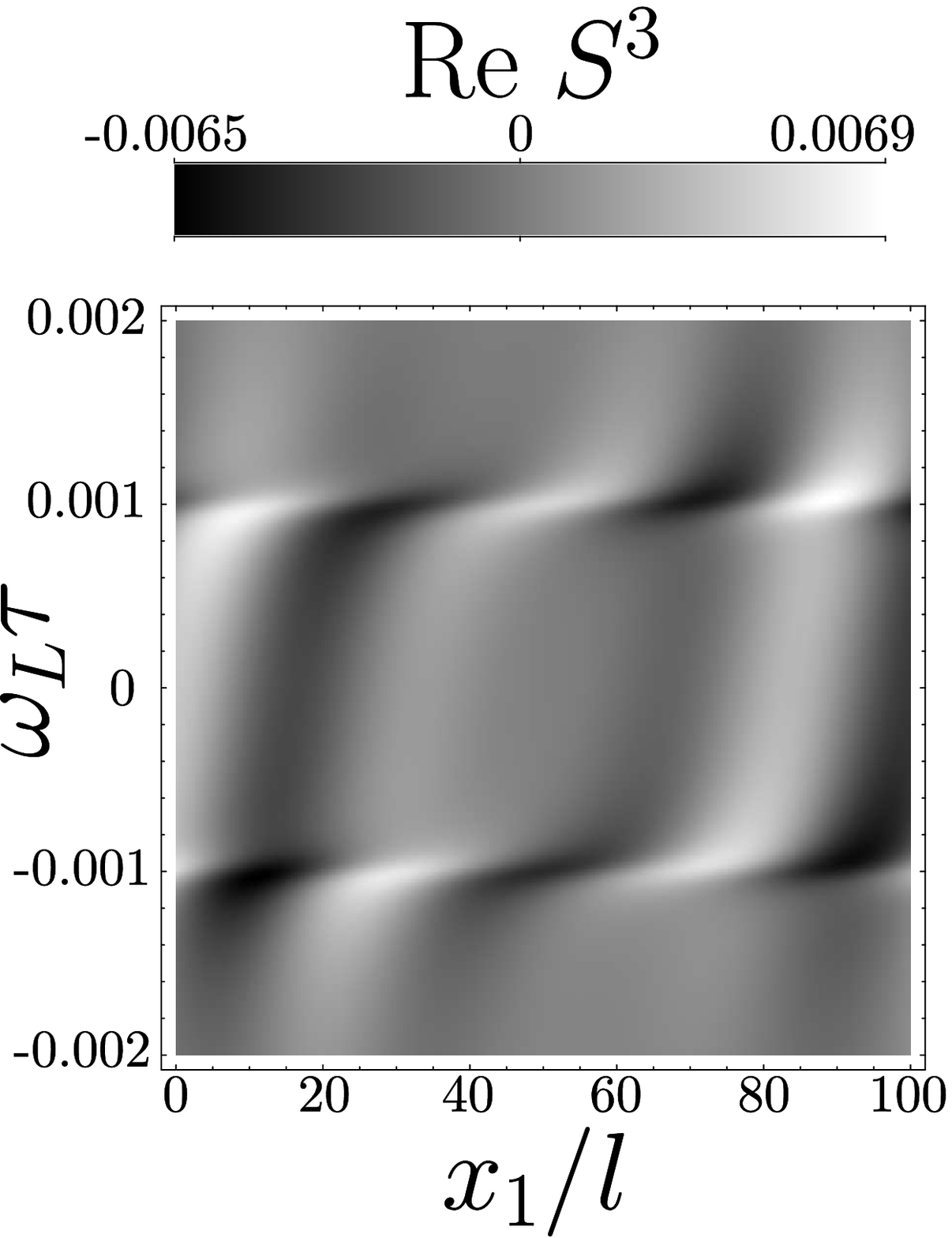}
  \includegraphics[width = 0.22 \textwidth]{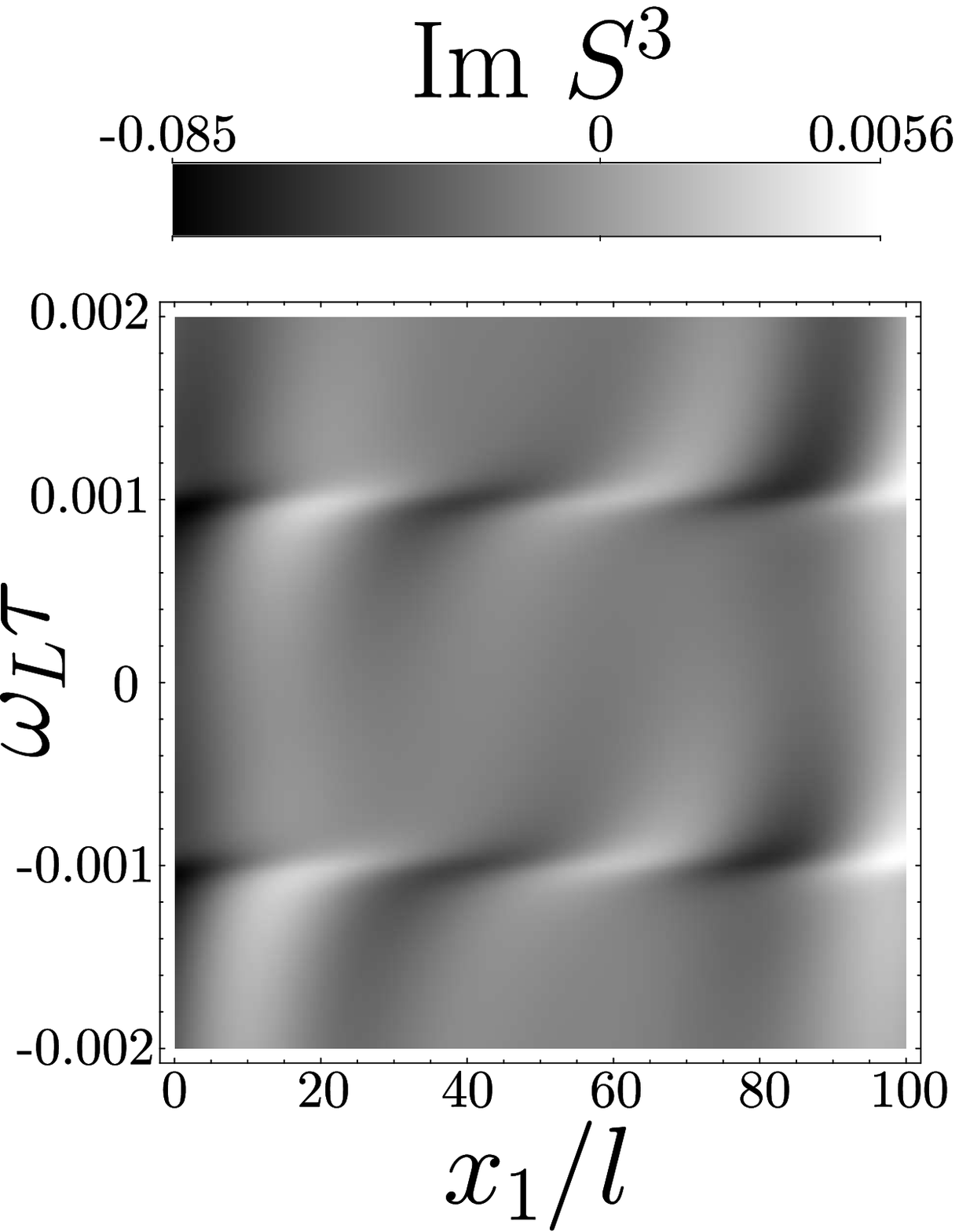}
  \caption{Polarization in the EDSR geometry $\mathbf E, \mathbf B ||
    \mathbf y$ for case shown in Fig.~\ref{fig:sample}(a) with $
    \mathbf S_b =  0.02 \mu \mathrm m^{-2}$. Upper panel: $\mathrm{Im}
    \scmd{S}^3(x=20 l)$ (black) and $\mathrm{Im} \scmd{S}^1(x=20 l)$ (grey
    curve) are shown as a function of $\omega_L \tau$.  Resonance is
    seen at $\pm \omega_L \tau = \omega \tau = 10^{-3}$.  Parameters
    of the $[001]$-grown QW: $\xi_\beta= - 0.08$, $\xi_\alpha= 0.1$, $L=100
    l$. Lower Panel: Density plot of $\mathrm{Re} \scmd{S}^3(x)$,
    $\mathrm{Im} \scmd{S}^3(x)$ as a function of $x$ and $\omega_L
    \tau/\xi_\alpha^2$ for the same parameters.}
  \label{fig:polarization-edsr}
\end{figure}

\section{EDSR and driven spin  helix}\label{sec:edsr-driven-spin}
We now focus on electric-dipole-induced spin resonance (EDSR)
\cite{Rashba1991, Bell1962,Rashba2003, Kato2004, Schulte2005}$^,$\cite{Duckheim2006,
  Wilamowski2007,Duckheim2007,Meier2007} in the finite Hall bar
geometry. We calculate the spin polarization $\boldsymbol{\scmd{S}}$
due to a simultaneous effect of ac electric field and dc magnetic
field $\mathbf b_0$, both along the channel. The directions of the
fields are chosen in such a way so that the internal field
$\boldsymbol{\Omega} (e \mathbf E(\omega) \tau )$ and $\mathbf b_0$
are perpendicular. \footnote{We also studied the Hanle geometry when
  $\boldsymbol{\Omega} (e \mathbf E(\omega) \tau ) || \mathbf b_0$;
  however, the effect of $b_0$ becomes appreciable only for fields as
  large as $b_0 \approx \tau^{-1}$ in that case. }  This geometry is
suitable for an observation of \textit{electrically} driven Rabi
oscillations of the spin polarization between the directions along and
opposite to $\mathbf b_0$.\cite{Rashba2003,Duckheim2006,Duckheim2007}

We focus on case (a) in Fig.~\ref{fig:sample}.  The magnetic field
$\mathbf {b}_0$ leads to an equilibrium polarization (Pauli
paramagnetism) $\boldsymbol{\scmd{S}}_{b_0} \propto \mathbf{\hat{x}}_2
\omega_L = 2 \mathbf{b}_0 $ in the longitudinal direction of the
channel. In addition, the polarization in the bulk of the sample
(transverse to $\mathbf b_0$) is modified.  In the geometry of
Fig.~\ref{fig:sample} (a) with $\mathbf b_0|| \mathbf e_{x_2}$, we
find for the bulk polarization
\begin{align}
  \label{eq:rho-in-edsr}
  \boldsymbol{\scmd{S}}_{\infty} &= \left(
  \begin{array}{c}
     (\omega_L^2 + \Gamma_-  (\Gamma_+ + \Gamma_- - i \omega) )\\ 0 \\ -i \omega \omega_L
  \end{array}
\right) \\ \notag &\times \frac{S_b  }{\omega_L^2 - \omega^2 - i \omega 
  (\Gamma_+ + 2 \Gamma_-) + \Gamma_-^2 + \Gamma_+ \Gamma_-} ,
\end{align}
where $\Gamma_{\pm} = 2 p_F^2 \tau (\alpha \pm \beta)^2$. In the
absence of the magnetic field, i.e., for $\omega_L=0$,
Eq.~(\ref{eq:rho-in-edsr}) reduces to
Eq.~(\ref{eq:rho-in}). Additionally, the characteristic modes change
due to $\mathbf b_0$. The wave numbers $\theta$ are determined by the
requirement of vanishing eigenvalues
\begin{align}
  \label{eq:evals-edsr}
  \frac{1}{2} \Gamma_++ \Gamma_- - D \theta ^2- i \omega \mp
  \frac{1}{2} \sqrt{\Gamma_+^2-4 \left(\omega _{L} - \theta
      C_-\right)^2}=0
\end{align}
of the differential operator $L(\theta)$ defined by
Eqs.~~(\ref{eq:rho-pde}),(\ref{eq:DE-operator}).

We focus on the case of $\alpha \approx -\beta$.  Expanding to first
order in 
$\Gamma_+ 
/\left(\omega _{L} - \theta C_-\right)\ll 1$, one
finds
\begin{align}
  \label{eq:theta-edsr}
  \theta_{1,2 } &= \mp l^{-1} \sqrt{2 \tau \left(\Gamma_+-i \omega
    \right)} \notag \, , \\ 
  \theta_{3,4 } &= \frac{
      i C_-  \mp \sqrt{2} \sqrt{D \left[\Gamma_+-2 i \left(\omega
              +\omega _{\text{L}}\right)\right]} }{2
      D} \notag  \, , \\ 
  \theta_{5,6 } &= \frac{
      - i C_-  \mp \sqrt{2} \sqrt{D \left[\Gamma_+-2 i \left(\omega
              -\omega _{\text{L}}\right)\right]} }{2
      D} \, .
\end{align}
At resonance, i.e., for $\omega_L = \omega$, the wave numbers
$\theta_{5,6}$ in Eq.~(\ref{eq:theta-edsr}) become purely imaginary
because $\Gamma_+ = 0$ for $\alpha = -\beta$. The modes $s_{5,6}$ are
thus completely undamped oscillations of the spin density with wave
length $\lambda^-_{SO} = 1/2 m (\beta - \alpha)$
[cf. Fig~\ref{fig:mode}]. Note that in the considered case of $\alpha
= -\beta$ the Hamiltonian commutes with the longitudinal spin (
$[H,\sigma^2]=0$), i.e., the $U(1)$-symmetry described in
Ref.~\onlinecite{Schliemann2003} remains intact; however, the
$SU(2)$-symmetry used in Ref.~\onlinecite{Bernevig2006} to demonstrate
the existence of the persistent spin helix is broken in the presence
of $\mathbf b_0$.

Figure~\ref{fig:polarization-edsr} shows a profile of the spin
polarization under EDSR conditions. At resonance ($\omega_L=\pm
\omega$), the overall amplitude of the out-of-plane polarization is
enhanced. This enhancement becomes particularly strong for $\Gamma_+
\approx 0$ occurring at $\alpha = -\beta$.

Solving the diffusion equation [Eq.~(\ref{eq:rho-pde})] to first order
in $\Gamma_+$ for the case $\alpha \approx - \beta$, $\omega \approx
+\omega_L$, we obtain the following expression for the spin density
close to resonance~($\omega \approx +\omega_L$)
\begin{align}
  \label{eq:rho-edsr}
 \boldsymbol{\scmd{S}} (r_1) & \approx \boldsymbol{\scmd{S}}_{\infty} +
\frac{
\left(S_{b}^{1} - \scmd{S}_{\infty}^{1}-i\scmd{S}_{\infty}^{3} \right) (\xi_\beta - \xi_\alpha)}{\sinh\left(L R /l\right) R} \left(
\begin{array}{c}
  i\\
  0\\
  1
\end{array} \right) \notag \\
& \times \Bigg[e^{-i r_1/ \lambda_{SO}^-} \cosh\left(R \left(L-r_1 \right)/l
\right)  \\ & \quad-e^{i(L-r_1)/ \lambda_{SO}^-} \cosh\left(R
  r_1/l\right)\Bigg],  \notag
\end{align}
where $R = \sqrt{\tau \Gamma_{+}-2i(\omega -
  \omega_{L})\tau}$. Equation (\ref{eq:rho-edsr}) describes a spin
density wave along the transverse direction of the Hall bar with wave
length $\lambda_{SO}^-$ and an amplitude proportional to
$1/\sinh\left(L R /l\right) R$. We discuss this result in more detail
below.  Inserting Eq.~(\ref{eq:rho-in-edsr}) for
$\boldsymbol{\scmd{S}}_{\infty}$ into Eq.~(\ref{eq:rho-edsr}) ,
setting $\omega=\omega_L$, and expanding the hyperbolic functions in
Eq.~(\ref{eq:rho-edsr}) for a narrow channel with width $L \ll
\lambda_{SO}^+= 1/2 m (\alpha + \beta)$, one obtains the dominant
$\alpha$-dependence of $\boldsymbol{\scmd{S}}$ around the $\alpha
\approx - \beta$ point as
\begin{align}
  \label{eq:rho-edsr-expl}
\scmd{S}^3(r_1) & 
\approx K
  \frac{\alpha+\beta}{(\xi_{\alpha} + \xi_{\beta})^{2}+2\tau
      \Gamma^{\mathrm{res}}} \notag \\ 
& \times e^{-i r_{1}/\lambda_{SO}^{-}}
\left[e^{iL/\lambda_{SO}^{-}}-1\right] ,
\end{align}
where $K =[-i\omega/\Gamma_-](-2\nu eE\tau)
(\xi_{\beta}-\xi_{\alpha})l/L $ depends only on the combination $\beta
- \alpha$. Here, we introduced a phenomenological linewidth
$\Gamma^{\mathrm{res}} = \Gamma^{\mathrm{res}}_y + 2
\Gamma^{\mathrm{res}}_x + \mathcal O ((\omega \tau)^2)$ to model the
regularization of the amplitude of $\scmd{S}^3$ at $\alpha + \beta =
0$, which for $\Gamma^{\mathrm{res}} =0 $ would diverge as $1/(\alpha
+ \beta)$. For $\alpha + \beta = 0$, the relaxation mechanisms due to
linear intrinsic SOIs, which are dominant for generic $\alpha \neq \pm
\beta$, are ineffective, and finite spin relaxation rates
$\Gamma^{\mathrm{res}}_x$ and $\Gamma^{\mathrm{res}}_y$ of the $x_1$
and $x_2$ spin components, respectively, are due to an extrinsic or
cubic Dresselhaus SOI.

Equation (\ref{eq:rho-edsr-expl}) describes a spin density wave
$\scmd{S}^3(r_1)$ at frequency $\omega$ with a spatial profile of the
form $e^{-i r_{1}/\lambda_{SO}^{-}}$. The real and imaginary parts of
$\scmd{S}^3$ have stationary nodes separated by the shortest of the
two SO lengths, i.e., $\lambda_{SO}^-$.  In addition, the spin profile
is subject to a quantization condition: $\boldsymbol{\scmd{S}}$ is
proportional to a factor $1-e^{i L/\lambda_{SO}^-}$, which vanishes
for $L= 2 \pi N \lambda_{SO}^-$ (with $N$ being an integer) and
becomes maximal for $L=(2N+1)\pi \lambda_{SO}^-$. The profile
described by Eq.~(\ref{eq:rho-edsr-expl}) arises due to an excitation
of the spin helix modes $s_{5,6}$ under the EDSR conditions.  The
spatial oscillations of these modes have the same \lq\lq magic\rq\rq\/
wave number $\theta = 1/\lambda_{SO}^-$ as the static persistent spin
helix.\cite{Schliemann2003,Bernevig2006} However, whereas the
persistent spin helix is time-independent, the spin profiles in
Eq.~(\ref{eq:rho-edsr}),(\ref{eq:rho-edsr-expl}) oscillate also in
time at each point $r_1$ with the frequency $\omega_0$ of the applied
electric field. [The explicit time-dependence, e.g., $\scmd{S}(r_1,t)
\propto \sin \left(r_1/\lambda_{SO}^- + \omega_0 t \right) $ for $L
=(2N+1)\pi \lambda_{SO}^- $ and for $\mathbf E(t) = \mathbf E_0
\cos(\omega_0 t)$, is obtained by inverse Fourier transform of
Eqs.~(\ref{eq:rho-edsr}),(\ref{eq:rho-edsr-expl}).]. This driven spin
helix is a generalization of a static spin helix structure to the
time-dependent case.

 
Spatial quantization due to the Hall-bar boundaries, moreover, leads
to further enhancement of the amplitude of the spin helix modes in the
EDSR regime. The amplitude $(\alpha+\beta) / \left[(\xi_{\alpha} +
  \xi_{\beta})^{2}+2\tau \Gamma^{\mathrm{res}}\right]$ is infinite for
$\alpha = -\beta$ in a model with strictly linear SOI, i.e. for
$\Gamma^{\mathrm{res}}= 0 $, but is regularized by the next-to-leading
order effects due to cubic Dresselhaus and extrinsic SOIs, giving rise
to a finite linewidth $\Gamma^{\mathrm{res}}$.  \footnote{A similar
  regularization from effects outside the linear model occurs for the
  lifetime of the static spin helix\cite{Bernevig2006}. See
  Ref.~\onlinecite{Stanescu2007}.} Such an enhancement of the amplitude
of the driven spin helix close to the $\alpha=-\beta$ point in relatively
narrow QWs may be observable, e.g., by optical
techniques.\cite{Kato2004b,Sih2005}

\section{Conclusions}

In conclusion, we have described several signatures of electrically
induced spin polarization and the spin-Hall effect due to {\it linear}
spin-orbit interactions. We have shown that the spin-Hall effect and
edge spin accumulation-- while being absent for dc electric fields--
becomes finite for time-dependent electric fields. In particular, we
have found that boundary effects can extend over the whole sample due
to driven spin helix modes for the case of the linear Rashba and
Dresselhaus spin-orbit interaction being of equal strengths. The
amplitude of these helix modes as a function of the spin-orbit
interaction strengths is strongly enhanced due to due spatial
quantization under the conditions of electric-dipole-induced spin
resonance.

\section*{ACKNOWLEDGMENTS}

M.D. and D.L. acknowledge financial support from the Swiss NF and the
NCCR Nanoscience Basel. D.L.M. acknowledges support from the Basel QC2
visitor program and 
NSF-DMR-0908029.
\appendix

\section{Spin diffusion equation}\label{sec:spin-diff-equat}

We start from the impurity averaged Kubo formula (for $E_F \tau \gg
1$) for the spin density
\begin{align}
  \label{eq:app-polarization-diagram}
\scmd{S}^i(\mathbf r) &= \left[
\begin{minipage}{.30\linewidth}
\includegraphics[width = 1.0 \textwidth]{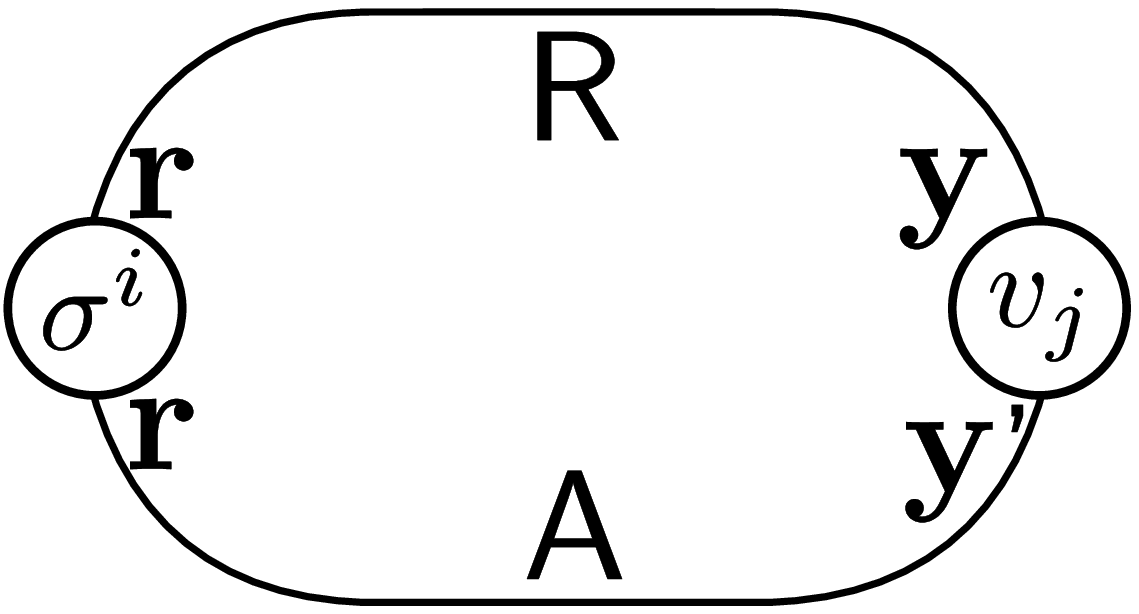}   
\end{minipage}
+ 
\begin{minipage}{.30\linewidth}
\includegraphics[width = 1.0 \textwidth]{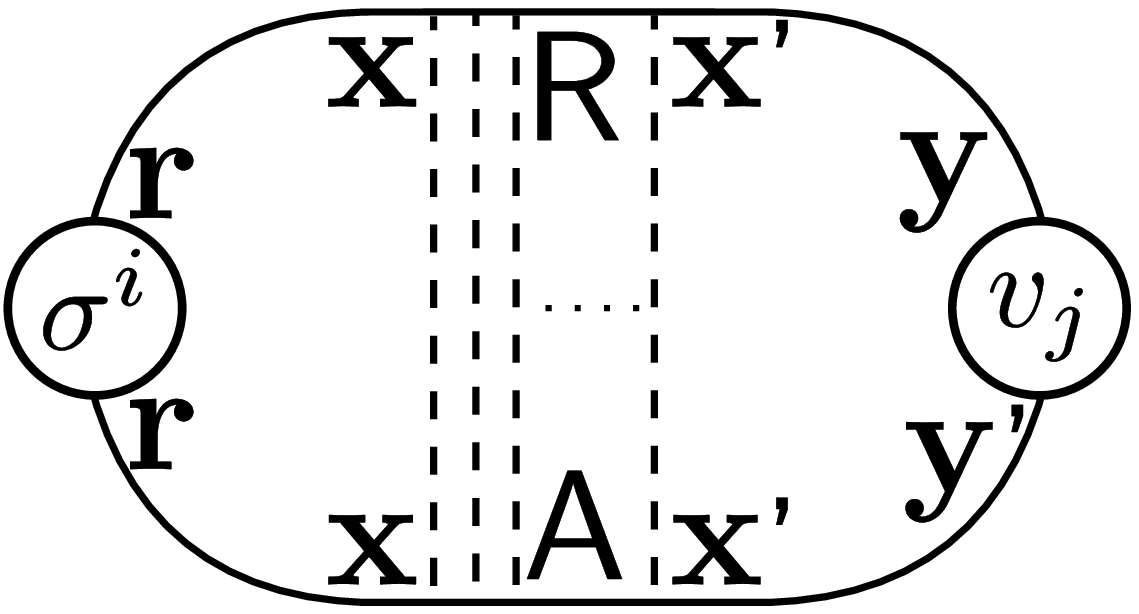}   
\end{minipage} \right] \frac{e}{2 \pi} E_j(\omega) \notag \\ &=
\frac{e}{2\pi}\int
d^{2}x'\Big[\delta^{i\nu}\delta(\mathbf{r}-\mathbf{x}') \notag
\\ &+\int
d^{2}x 2m\tau
X^{i\mu}(\mathbf{r},\mathbf{x})
D^{\mu\nu}(\mathbf{x},\mathbf{x}')\Big]\gamma^{\nu}(\mathbf{x}') \, ,
\end{align}
where solid lines denote impurity averaged Green's functions $G^{R/A}_E$
and dashed lines
denote 
correlators of impurity potential.
\cite{Duckheim2006,Akkermans2007, Chalaev2009}  The first term of
Eq.~(\ref{eq:app-polarization-diagram}) is the \lq\lq bubble\rq\rq\/ diagram
$\gamma^{\nu}(\mathbf{r})=\mathrm{tr}\left\{
  \langle\mathbf{r}|\sigma^{\nu}G^{R}_{E_F+\omega}\hat{v}_{j}G^{A}_{E_F}|\mathbf{r}\rangle\right\}
E_{j}(\omega)$, i.e., 
a spin response to the electric field in the absence of vertex corrections.
The latter are decribed
the diffuson $D^{\mu \nu} (\mathbf r,\mathbf x)$,
which
is defined by the integral equation
\begin{align}
\label{eq:diffuson-integral}
  D^{\mu \nu} (\mathbf r,\mathbf x') = \frac{\delta^{\mu
    \nu}\delta(\mathbf{r}-\mathbf{x'})}{2m\tau} + \int d^{2}y
  X^{\mu \rho}(\mathbf{r},\mathbf{y})D^{\rho
    \nu}(\mathbf{y},\mathbf{x'}) ,
\end{align}
where $X^{\mu\nu}$ is given 
by Eq.~(\ref{eq:spin-spin-diagram}). Iterating
Eq.~(\ref{eq:app-polarization-diagram}) once 
with the help of
Eq.~(\ref{eq:diffuson-integral}), we find
\begin{align}
  \label{eq:rho-integral-eq}
\scmd{S}^{i}(\mathbf{r})=\frac{e}{2\pi}(2m\tau)\int
  d^{2}x'D^{i\nu}(\mathbf{r},\mathbf{x'})\gamma^{\nu}(\mathbf{x'}) .
\end{align}
Multiplying Eq.~(\ref{eq:diffuson-integral}) by
$\frac{e}{2\pi}2m\tau\gamma^{\nu}(\mathbf{x'})$ and integrating over
$\mathbf x'$, we obtain the integral equation for the spin density
\begin{align}
  \label{eq:rho-integral-eq2}
\scmd{S}^{i}(\mathbf{r})=\frac{e}{2\pi}\gamma^{i}(\mathbf{r})+\int
  d^{2}x\, X^{i\nu}(\mathbf{r},\mathbf{x})\scmd{S}^{\nu}(\mathbf{x}) \, ,
\end{align}
which can be further simplified by partially evaluating the \lq\lq bubble\lq\lq\/
term $\gamma^{i}(\mathbf{r})$ in Eq.~(\ref{eq:rho-integral-eq2}). We
define the spin-momentum 
correlation functions
\cite{Duckheim2006}
\begin{align}
  \label{eq:y-diagram}
  Y^{\eta j}(\mathbf r) &=  \int \frac{ d^2x}{2m\tau}  \mathrm{tr}
  \left\{ \sigma^{\eta}G^{R}(\mathbf{r},\mathbf{x}) \frac{-i
      \partial}{m \partial
      x_{j}}G^{A}(\mathbf{x},\mathbf{r})\right\}  \\ 
  Y_{\mathrm{b}}^{\eta j} &= -\frac{\Omega_{\eta j}}{1-i \omega \tau}
\approx -(1 + i\omega \tau)  \Omega_{\eta j}  ,\label{eq:y-diagram-bk}
\end{align}
where Eq.~(\ref{eq:y-diagram-bk}) is obtained by
evaluating Eq.~(\ref{eq:y-diagram}) using the bulk Green's functions of
an infinite sample. Inserting the definition of the velocity operator
$\hat v_j = \frac{\hat p_{j}}{m}+\Omega_{kj}\sigma^{k}$, we obtain
$(e/2\pi) \gamma^i(\mathbf r) = - \int d^2 x X^{ik}(\mathbf r, \mathbf
x) S^k_b + 2 m \tau (e/2\pi) Y^{ij}(\mathbf r) E_j$. We can rewrite
Eq.~(\ref{eq:rho-integral-eq2}) as
\begin{align}
\label{eq:rho-integral-eq-part-ev}
  \scmd{S}^{i}(\mathbf{r}) - S_{b}^{i} & = i \omega \tau
  S_{b}^{i} +   \int d^{2}x\,
  X^{i j}(\mathbf{r},\mathbf{x}) (\scmd{S}^{j}(\mathbf{x}) -
  S_{b}^{j}) \notag \\ & + \left[Y(\mathbf{r})-Y_{\mathrm{b}}\right]^{ij}
  2\nu eE_{j}\tau  \, . 
\end{align}
From
now on, we treat the regions close to the boundary and in the bulk
separately.
In the bulk, one obviously has
$\left[Y(\mathbf{r})-Y_{\mathrm{b}}\right] = 0$ and arrives thus at
Eq.~(\ref{eq:rho-integral-eq3}).
At the boundary, the Green's functions
$G^{R/A}_0=G_{\mathrm{b},0}^{R/A}-G_{\mathrm{b},0}^{\ast R/A}$
constructed in Sec.~\ref{sec:boundary-conditions} have to be used to
evaluate $\gamma^i(\mathbf r)$, $Y^{\mu\mu}(\mathbf r)$.  Neglecting terms
oscillating with a period of $1/p_F$, as described in
Sec.~\ref{sec:boundary-conditions-1}, one finds that $ Y^{i j}(\mathbf
r) E_j = Y^{ij}_{\mathrm{b}} E_j $ to linear order in the SOI. 
Therefore,
the last term in Eq.~(\ref{eq:rho-integral-eq-part-ev}) vanishes.
Consequently,
Eq.~(\ref{eq:rho-integral-eq-part-ev}) turns into
Eq.~(\ref{eq:rho-integral-eq3}) and can be used for the derivation of
both the bulk diffusion equation and the boundary conditions.


\section{Boundary conditions}\label{sec:boundary-conditions-2}

For the coefficients $B$ and $C$ in Eq.~(\ref{eq:boundary-cond})
describing a boundary with normal vector $\mathbf{\hat n}$, we found
\begin{align}
  \label{eq:boundary-coefficients-2}
  \delta B^{\mu \nu}(\mathbf{\hat n}) \equiv [B - \mathbb 1]^{\mu \nu} & =
  -\frac{2}{\pi} 2 p_F \tau
  \boldsymbol{\Omega}^m(\mathbf n) \epsilon_{m \nu \mu} , \\
    \label{eq:boundary-coefficients-2-diff}
  C_j^{\mu \nu}(\mathbf{\hat n}) & = \frac{2}{\pi} \delta^{\mu \nu} l
  \mathbf{\hat{n}}\cdot \mathbf e_j,
\end{align}
where we neglected terms proportional to $\omega \tau \ll 1$. We
define
a
 $6 \times 6$ matrix
\begin{widetext}
\begin{align}
  \label{eq:BC-matrix-form}
  M =
  \begin{scriptsize}
    \left(\begin{array}{cccc} \left(\delta B(\mathbf{\hat n})
          +\theta_{1} C(\mathbf{\hat n})\right) \mathbf{s}_{1,0}
        e^{\theta_1 r}\Bigg|_{r=0} & \left(\delta B(\mathbf{\hat n})
          +\theta_{2} C(\mathbf{\hat n})\right) \mathbf{s}_{2,0}
        e^{\theta_2 r}\Bigg|_{r=0} & \dots & \left(\delta
          B(\mathbf{\hat n}) +\theta_{6} C(\mathbf{\hat n})\right)
        \mathbf{s}_{6,0} e^{\theta_6 r}\Bigg|_{r=0}\\
        \left(\delta B(-\mathbf{\hat n}) +\theta_{1} C(\mathbf{-\hat
            n})\right) \mathbf{s}_{1,0} e^{\theta_1 r}\Bigg|_{r=L} &
        \left(\delta B(\mathbf{-\hat n}) +\theta_{2} C(\mathbf{-\hat
            n})\right) \mathbf{s}_{2,0} e^{\theta_2 r}\Bigg|_{r=L} &
        \dots & \left(\delta B(\mathbf{-\hat n}) +\theta_{6}
          C(\mathbf{-\hat n})\right)
        \mathbf{s}_{6,0} e^{\theta_6 r}\Bigg|_{r=L}\\
    \end{array}\right)    
  \end{scriptsize}
\end{align}
\end{widetext}
and a vector $\mathbf A = (\mathbf A_0, \mathbf A_L)$, where $\mathbf
A_{0,L} = \delta B(\mathbf{\pm \hat n})
(\boldsymbol{\scmd{S}}_{\infty} - \mathbf S_b) $. Inserting the
general solution $\boldsymbol{\scmd{S}} =
\boldsymbol{\scmd{S}}_{\infty} + c_k \mathbf s_{0,k} e^{\theta_k r_1}
$ into Eq.~(\ref{eq:boundary-cond}), the BCs can be rewritten as $M
\mathbf c = -\mathbf A$.


\end{document}